\documentstyle[12pt]{article}

\begin{document}
\mbox{ }
\rightline{UCT-TP-255/99}\\
\rightline{March 1999}\\
\rightline{Revised February 2000}\\
\vspace{3.5cm}
\begin{center}
{\Large \bf Is there evidence for dimension-two corrections in \\
[.3cm] QCD two-point functions ?}
\footnote{Work supported
in part by the Volkswagen Foundation}\\
\vspace{.5cm}
{\bf C. A. Dominguez$^{(a)}$, and K. Schilcher$^{(b)}$}\\[.5cm]
$^{(a)}$Institute of Theoretical Physics and Astrophysics\\
University of Cape Town, Rondebosch 7700, South Africa\\[.5cm]
$^{(b)}$Institut f\"{u}r Physik, Johannes Gutenberg-Universit\"{a}t\\
Staudingerweg 7, D-55099 Mainz, Germany
\end{center}
\vspace{.5cm}
\begin{abstract}
\noindent
The ALEPH data on the (non-strange) vector and axial-vector spectral 
functions, extracted from tau-lepton decays, is used in order to search for 
evidence for a dimension-two contribution, $C_{2 \; V,A}$, to the Operator 
Product Expansion (other than $d=2$ quark mass terms).
This is done by means of a dimension-two Finite Energy 
Sum Rule, which relates QCD to the experimental hadronic information. The 
average $C_{2} \equiv (C_{2V} + C_{2A})/2$ is remarkably stable against 
variations in the
continuum threshold, but depends rather strongly on $\Lambda_{QCD}$. Given
the current wide spread in the values of $\Lambda_{QCD}$, as extracted from
different experiments, we  conservatively conclude from our
analysis that $C_{2}$ is
consistent with zero. 
\end{abstract}
\newpage
\setlength{\baselineskip}{1.5\baselineskip}
\noindent

The Operator Product Expansion (OPE), extended beyond perturbation theory,
is one of the pillars of the successful QCD sum rule method 
used extensively to link QCD and hadronic physics \cite{SVZ}. In 
analyzing e.g. two-point functions,
one calculates QCD perturbative contributions up to a desired order in
the running strong coupling, and then includes non-perturbative effects 
parametrized
in terms of a series of vacuum to vacuum matrix elements of local, gauge
invariant operators built from the quark and gluon fields entering the
QCD Lagrangian. These so called vacuum condensates encompass the long
distance dynamics. They are multiplied by Wilson coefficients, calculable
in perturbation theory, which contain the short distance information. For
a given dimension, these terms fall off as powers of the (Euclidean)
momentum transfer $q^{2}$ ($q^{2} < 0$). The lowest naive dimension in QCD
is $d=4$, corresponding to the gluon condensate and to the product of quark 
masses
and the quark condensate. In the standard OPE approach there are no
dimension-two corrections, other than the well known quark mass insertion 
contributions of the form $m_{q}^{2}/q^{2}$. Except possibly in the
strange-quark sector, these terms can be safely neglected for
light quark current correlators. On the
other hand, some specific dynamical
mechanisms have been suggested as potential sources of $d=2$ corrections 
to the OPE, e.g. renormalons \cite{REN} or a tachyonic gluon mass 
\cite{GM}. If present, these dimension-two corrections are not expected to
have much impact on most of the existing phenomenological predictions from 
QCD sum rules, as the level of precision is seldom better than 10-20 \%.
However, they may have a non-negligible impact on the extraction of the
QCD strong coupling at the tau-lepton mass scale, $\alpha_s(M_{\tau}^2)$.
In fact, current  claims from this precision determination \cite{BR}
have been questioned \cite{AL} on the grounds that they rely on the
assumption of no $d=2$ corrections to the OPE (other than
quark mass insertions). Attempts have been made to determine 
phenomenologically the size of potential dimension-two terms using 
information on (i) the vector spectral function as obtained from 
$e^{+} e^{-}$ data 
\cite{NAR1}, and  (ii) the vector and the axial-vector spectral
functions extracted from tau-lepton hadronic decays \cite{CAD1}, as 
measured by
the ARGUS collaboration \cite{ARGUS}. The analysis based on tau decay
data has the
advantage of relying on two independent quantities ($d=2$ corrections are
expected to be chiral-symmetric), which can be constrained further by
using theoretical information such as the first Weinberg sum rule.
The conclusion from this determination \cite{CAD1} was that the ARGUS
data supported a non-zero dimension-two term in the OPE, which was
consistent with a  dependence on the QCD scale of the form:
$C_{2} \propto \Lambda_{QCD}^{2}$. With the advent of the ALEPH precision 
data \cite{ALEPH} on semileptonic tau decays it is now possible to
analyze a variety of theoretical issues involving the vector and
axial-vector spectral functions. One such issue is, precisely, the possible
presence of a dimension-two contribution in the OPE, which is the
subject of this note.
These spectral functions are
related to the discontinuities in the complex energy plane of the
two-point functions involving the vector and axial-vector currents
\begin{eqnarray*}
 \Pi_{\mu \nu}^{VV} (q^2) = i \; \int \; d^4 \; x \; e^{i q x} \; \;
  <0|T(V_{\mu}(x) \; \; V_{\nu}^{\dagger}(0))|0> \; 
\end{eqnarray*}
\begin{equation}
  = \; (- g_{\mu \nu} \; q^{2} + q_{\mu} q_{\nu}) \; \Pi_{V} (q^{2}) \; ,
\end{equation}
\begin{eqnarray*}
 \Pi_{\mu \nu}^{AA} (q^2) = i \; \int \; d^4 \; x \; e^{i q x} \; \;
  <0|T(A_{\mu}(x) \; \; A_{\nu}^{\dagger} (0) )|0> \; 
\end{eqnarray*}
\begin{equation}
  = \; (- g_{\mu \nu} \; q^{2} + q_{\mu} q_{\nu}) \; \Pi_{A} (q^{2})
  - q_{\mu}  q_{\nu} \; \Pi_{0} (q^{2})  \; ,
\end{equation}
where $V_{\mu} = :(\bar{u}  \gamma_{\mu}  u-\bar{d}  
\gamma_{\mu}  d):/2$, and
$A_{\mu} = :(\bar{u}  \gamma_{\mu}  \gamma_{5} u- \bar{d}
\gamma_{\mu} \gamma_{5} d):/2$. Considering these (charge neutral)
currents implies the normalization $Im \;\Pi_{V} = Im \;\Pi_{A} =
1/8 \pi$, at leading order in perturbative QCD.
In order to determine the size of a potential dimension-two contribution
to the OPE we consider the following Finite Energy Sum Rule (FESR)
\cite{FESR}
\begin{equation}
I_{0\;V,A} \;  \equiv \; \frac{8 \pi^{2}}{s_{0}} 
\int_{0}^{s_{0}} \; 
 \rho_{\;V,A} (s) \; ds \;
   = \; \frac{C_{2\;V,A}}{s_{0}} \; + F_{2}(s_{0}) \;,
\end{equation}
where $C_{2 \; V,A}$ is the potential $d=2$ term, $s_{0}$ is the continuum 
threshold, and $F_{2}(s_{0})$ is the
radiative correction, identical in the vector and axial-vector channels,
which is obtained after a straightforward integration of the perturbative
QCD results of \cite{CHET}; to four-loop order it is given by
\begin{eqnarray*}
F_{2}(s_{0}) = 1 + \frac{\alpha^{(1)}_{s}(s_{0})}{\pi}
+ \Biggl ( \frac{\alpha^{(1)}_{s}(s_{0})}{\pi} \Biggr )^{2}
\;\Biggl (F_{3} - \frac{\beta_{2}}{\beta_{1}} {\rm ln} L - 
\frac{\beta_{1}}{2}
\Biggr )
\end{eqnarray*}
\begin{equation}
+ \Biggl ( \frac{\alpha^{(1)}_{s}(s_{0})}{\pi}\Biggr )^{3}
\Biggl [ \frac {\beta_{2}^{2}}{\beta_{1}^{2}} ( {\rm ln} ^{2} L - 
{\rm ln} L -1)
+\frac{\beta_{3}}{\beta_{1}} - 2 \Biggl ( F_{3} - \frac{\beta_{1}}{2} 
\Biggr ) \frac{\beta_{2}}{\beta{1}} {\rm ln} L + F_{4} - F_{3} \beta_{1}-
\frac{\beta_{2}}{2} + \frac{\beta_{1}^{2}}{2} \Biggr ] \; ,
\end{equation}
with
\begin{equation}
\frac{\alpha^{(1)}_{s}(s_{0})}{\pi} \equiv \frac{- 2}{\beta_{1} L}\; ,
\end{equation}
where $L \equiv {\rm ln} (s_{0}/\Lambda^{2}_{QCD})$, and for three flavours:
$\beta_{1} = - 9/2$, $\beta_{2} = - 8$, $\beta_{3} = - 3863/192$,
$F_{3} = 1.6398$, $F_{4} = - 10.2839$. In writing Eq. (4) use has been made
of the result \cite{CHET}
\begin{eqnarray*}
\frac{\alpha^{(3)}_{s}(s_{0})}{\pi} =
\frac{\alpha^{(1)}_{s}(s_{0})}{\pi}
+ \Biggl (\frac{\alpha^{(1)}_{s}(s_{0})}{\pi}\Biggr )^{2}
\Biggl (\frac{- \beta_{2}}{\beta_{1}} {\rm ln} L \Biggr )
\end{eqnarray*}
\begin{equation}
+  \Biggl (\frac{\alpha^{(1)}_{s}(s_{0})}{\pi}\Biggr )^{3} 
\Biggl (\frac{\beta_{2}^{2}}{\beta_{1}^{2}} ( {\rm ln}^{2} L -
{\rm ln} L -1) + \frac{ \beta_{3}}{\beta_{1}} \Biggr ) \; ,
\end{equation}
and an expansion in powers of $\alpha^{(1)}_{s}$ is to be understood.
Alternatively, one may choose not to expand in such a way; in this case
the radiative correction becomes
\begin{eqnarray*}
F_{2}(s_{0}) = 1 + \frac{\alpha^{(3)}_{s}(s_{0})}{\pi}
+ \Biggl ( \frac{\alpha^{(3)}_{s}(s_{0})}{\pi} \Biggr )^{2}
\;\Biggl (F_{3}  - 
\frac{\beta_{1}}{2}
\Biggr )
\end{eqnarray*}
\begin{equation}
+ \Biggl ( \frac{\alpha^{(3)}_{s}(s_{0})}{\pi}\Biggr )^{3}
\Biggl ( 
  F_{4} - F_{3} \beta_{1}-
\frac{\beta_{2}}{2} + \frac{\beta_{1}^{2}}{2} \Biggr ) \; ,
\end{equation}
where $\alpha^{(3)}_{s}(s_{0})$ is given by Eq. (6). Numerically, these
two alternatives have a non-negligible impact on the final result for
$C_{2}$, as will be discussed later.\\
The quark mass insertion term, which contributes to this dimension-two
FESR is of the form
\begin{equation}
C_{2m} = -3 \frac{(\hat{m}_{u}^{2}+\hat{m}_{d}^{2})}
{(\frac{1}{2} {\rm ln} s_{0}/\Lambda^{2}_{QCD})^{-4/\beta_{1}}}
\end{equation}
Using current values of the up- and down-quark masses, this term is 
negligible.\\

It should be stressed at this point that the FESR are
ideally suited, in principle, to extract the values of power corrections of
a given dimensionality. Ignoring gluonic corrections to the condensates, the
FESR involving $\rho_{V,A}$ with kernel $s^{N}$ (N=0,1,2,...) project out 
only condensates of dimension $d=2,4,6,...$ .In other words, in the FESR of 
lowest dimension all condensates of $d=4,6,...$ decouple. This should be
contrasted with
e.g. Laplace or Gaussian sum rules which receive contributions from all
possible condensates. Since the numerical values of these power corrections 
are not well known, these other sum rules introduce an unnecessary additional
uncertainty. On the other hand, the fact that FESR tend to emphasize the high 
energy region, where the ALEPH data have larger errors, is of no importance
here, as
the uncertainty in $C_{2}$ turns out to be overwhelmingly dominated by
the uncertainty in $\Lambda_{QCD}$, and to a lesser extent, by the way
the perturbative expansion is organized. The experimental error in the
hadronic integral in Eq.(3) can be safely neglected.\\

We show now that the dimension-two terms obtained from Eq.(3), i.e.
\begin{equation}
C_{2\;V,A} = 8 \pi^{2}  \int_{0}^{s_{0}} \; 
 \rho_{\;V,A} (s) \; ds \;
   - \; \;\;s_{0}  F_{2}(s_{0}) \; ,
\end{equation}
are actually identical in the vector and axial-vector channels, provided
one takes the chiral limit. This result would follow trivially from e.g. the
first Weinberg sum rule, provided this sum rule would be saturated by the
data for $s_{0} < \infty$ (actually, $s_{0} < M_{\tau}^{2}$ in the case
of tau decay data). However, this is not the case, as discussed in
\cite{CHSR}. Instead, the data saturate much better the  modified sum rule
 \cite{CHSR}
\begin{equation}
\int_{0}^{s_{0}} \;      (1 - \frac{s}{s_{0}})\;
[ \rho_{V} (s)-  \rho_{A} (s)] \; ds \;= 0 \; ,
\end{equation}
 Here, $\rho_{A}$ already contains the pion pole, i.e.
\begin{equation}
\rho_{A}(s) = f_{\pi}^{2} \delta (s) + \rho_{A} (s) |_{RES} \;
\end{equation}
where $\rho_{A} (s) |_{RES}$ is the resonance part of the spectral function.
We make  use of Eq.(9) in Eq.(10), and invoke the dimension-four FESR
\begin{equation}
I_{1\;V,A} \;  \equiv \; \frac{8 \pi^{2}}{s_{0}^{2}} 
\int_{0}^{s_{0}} \; 
 \rho_{\;V,A} (s) \; s \; ds
   = \frac{ F_{4}(s_{0})}{2} \; -
     \; \frac{C_{4}<O_{4}>}{s_{0}^{2}} \;
\end{equation}
where both the radiative correction $F_{4}(s_{0})$, and the dimension-four
condensate (equal to the gluon condensate in the chiral limit), are
identical in the vector and axial-vector channel. One then obtains
\begin{equation}
\int_{0}^{s_{0}} \;      (1 - \frac{s}{s_{0}})\;
[ \rho_{V} (s)-  \rho_{A} (s)] ds \;= 0
= \frac{1}{8 \pi^{2}} (C_{2V} \; - C_{2A})\; ,
\end{equation}
which completes the proof.

We proceed now to use the ALEPH data in Eq.(9) in order to check for
any evidence for a dimension-two operator. 
The results of a numerical evaluation of the r.h.s. of Eq.(9), using the
expanded form of the QCD integral, Eq. (4), and a fit to the ALEPH data
as described in \cite{CHSR},
are plotted as a function of $s_{0}$ in
Fig.1, for $\Lambda_{QCD} = 300 \; \mbox{MeV}$. Curve (a) is the result in 
the vector channel, curve (b) in the axial-vector channel, and curve
(c) is the average between the two, i.e. $C_{2} = (C_{2V}+C_{2A})/2$.
It can be seen that the results for $C_{2V}$ and $C_{2A}$, considered 
individually, are rather  unstable against variations in $s_{0}$. This
behaviour is simple to understand. When integrating the data, the vector
integral at $s_{0} \simeq 1 \; \mbox{GeV}^{2}$ has already picked up the
contribution from the (narrow-width) rho-meson, while the axial-vector
spectral function is still relatively small there. The contribution from the
(broad-width) $a_{1}$-meson is important only for $s_{0} > 1.5 - 1.6
\; \mbox{GeV}^{2}$. For this reason, the hadronic integral approaches the
theoretical or QCD integral from above in the vector channel, and from below
in the axial-vector channel at $s_{0} < 1.5 \; \mbox{GeV}^{2}$. The 
expectation  $C_{2V} = C_{2A}$ is only true asymptotically.
However, making use of 
this constraint,  it is natural to consider instead the 
average value. This turns out to be remarkably stable, and allows for a 
precise determination of $C_{2}$. Perhaps this should not be surprising,
as the dimension-six four-quark condensate contributes with different
signs to the vector and axial-vector correlators, and there is a 
tendency to an overall cancellation between the sum of the gluon
and the four-quark condensates in $\Pi_{V}+\Pi_{A}$ (which, however,
enter the sum rule under consideration only via radiative corrections).
Nevertheless, there is a strong dependence
of $C_{2}$ on $\Lambda_{QCD}$ as discussed next. The size of the current
error bars in  $\Lambda_{QCD}$ makes it the dominant source of
uncertainty in the determination of $C_{2}$; in comparison, the small 
experimental uncertainty in the hadronic spectral functions plays a 
negligible role.

First, we have found a strong dependence of $C_{2}$ on the order at which 
the radiative correction $F_{2}$, Eq.(4),  is computed. 
As the right hand side of Eq.(9) is the (small) difference of two similar
numbers, it is not surprising that the transition from one-loop to two-loops 
of perturbative QCD in Eq.(4) leads to relatively large differences.
In other words, the result for $C_2$ obtained using Eq.(4) truncated to
one loop, i.e. using only the first term in that equation, differs
substantially from the result for $C_2$ if the truncation is to two-loops,
i.e. including the first two terms in Eq.(4).
Higher order corrections (three-, four-, and five-loops) show, however, the 
expected convergence, i.e. there is little difference between the values of
$C_2$ obtained after truncation at the three-loop,
four-loop, and five-loop level. In the latter case, Eq.(7) becomes
\begin{eqnarray*}
F_{2}(s_{0}) &=& 1 + \frac{\alpha^{(4)}_{s}(s_{0})}{\pi}
+ \Biggl ( \frac{\alpha^{(4)}_{s}(s_{0})}{\pi} \Biggr )^{2}
\;\Biggl (F_{3}  - 
\frac{\beta_{1}}{2}
\Biggr ) \\
&+& \Biggl ( \frac{\alpha^{(4)}_{s}(s_{0})}{\pi}\Biggr )^{3}
\Biggl ( 
  F_{4} - F_{3} \beta_{1}-
\frac{\beta_{2}}{2} + \frac{\beta_{1}^{2}}{2} \Biggr ) \; 
\end{eqnarray*}
\begin{equation}
+ \Biggl ( \frac{\alpha^{(4)}_{s}(s_{0})}{\pi}\Biggr )^{4}
\Biggl [ k_3 - \frac{3}{2} \beta_1 F_4 + \frac{\beta_1^2}{2} F_3 (3 -
\frac{\pi^2}{2}) - \frac{3}{4} \beta_1^3 - \beta_2 F_3
+ \frac{5}{4}  \beta_1 \beta_2 (1 - \frac{\pi^2}{6})
 - \frac{\beta_3}{2} \Biggr ] \; ,
\end{equation}

where $\alpha^{(4)}_{s}(s_{0})$ is given by \cite{B5}
\begin{eqnarray}
\frac{\alpha^{(4)}_{s}(s_{0})}{\pi} =
\frac{\alpha^{(3)}_{s}(s_{0})}{\pi}
- \Biggl (\frac{\alpha^{(1)}_{s}(s_{0})}{\pi}\Biggr )^{4}
\Biggl [\frac{\beta_{2}^3}{\beta_{1}^3} ({\rm ln}^3 L
-\frac{5}{2} {\rm ln}^{2} L - 2 {\rm ln} L + \frac{1}{2})
+ 3 \frac{\beta_2 \beta_3}{\beta_1^2} {\rm ln} L + \frac{b_3}{\beta_1}
\Biggr ] \; ,
\end{eqnarray}
and
\begin{equation}
b_3=\frac{1}{4^4} \Biggl[ \frac{149753}{6} + 3564 \zeta_3
-(\frac{1678361}{162} + \frac{6508}{27} \zeta_3 ) n_F
+ (\frac{50065}{162} + \frac{6472}{81} \zeta_3 ) n_F^2 +
 \frac{1093}{729}
n_F^3 \Biggr ] \; ,
\end{equation}
and $\zeta_3 = 1.202$.
While the QCD beta-function is known to
four loops \cite{B5}, the imaginary part of the vector (axial-vector)
correlator to five-loop order involves the unknown constant
$k_3$ in Eq. (14).
We have estimated this constant assuming a geometric series behaviour
for those constants not determined by the renormalization group,
i.e. $k_3 \simeq k_2^2/k_1 \simeq 25$, with $ k_1 \equiv F_3$, and
$k_2 \simeq F_4 + \pi^2 \beta_1^2/12$. This is in good qualitative
agreement with other estimates \cite{K3}.
In this case we find that the difference between the four-loop and the
five-loop results for $C_{2}$ is again quite small. 
Nonetheless, the four- and five-loop determinations (represented in Fig.2)
depend strongly 
on the QCD scale parameter $\Lambda _{QCD}$ (here $\Lambda _{QCD}
\equiv \Lambda _{\overline{MS}}(n_F=3)$).
As $\Lambda _{QCD}$ is varied
in the  range \cite{PDG} 
$300 \;\mbox{MeV} \leq \Lambda _{QCD}\leq  400 \;\mbox{MeV}$, the
extracted value of $C_{2}$ remains small and changes sign after crossing
zero for $\Lambda _{QCD} \simeq 330 - 360 \; \mbox{MeV}$, for typical values
of $s_{0}$ in the stability region ($s_{0} \simeq 1.5 - 2.7 
\; \mbox{GeV}^{2}$). In Fig. 3 we show the
dependence of $C_{2}$ on $\Lambda_{QCD}$ for a typical value $s_{0}
= 2 \;\mbox{GeV} ^{2}$ at the center of the stability region. Such a strong
dependence of $C_{2}$ on $\Lambda_{QCD}$ could be a welcome feature, if
for some theoretical reason one could rule out completly the presence of a 
dimension-two operator in QCD. One could then determine $\Lambda_{QCD}$
with great accuracy. Finally, the difference in the QCD integral after
expanding in powers of $\alpha^{(1)}_{s}$, as in Eq.(4), or not expanding them,
as in Eq.(7),
though small, does affect the value of $C_{2}$, as this is the result
of the subtraction of two very similar numbers. For $\Lambda_{QCD}$ as
determined by the PDG \cite{PDG},
i.e. $\Lambda_{QCD} = 389 \pm 35 \;\mbox{MeV}$,
we find $C_{2} = -(0.08 \pm 0.06)\; \mbox{GeV}^{2}$ if the expansion in
$\alpha^{(1)}_{s}$ is done as in Eq. (4), and
$C_{2} = - 0.05 \pm 0.05 \; \mbox{GeV}^{2}$ if one does not expand, 
as in Eq.(7).
Alternatively, we  consider the value of $\alpha_{s}$
extracted in \cite{ALEPH}, which implies $\Lambda_{QCD}= 367 \pm 40 \;
\mbox{MeV}$, for three flavours. This analysis was performed assuming 
$C_{2} =0$. In view of our results, this assumption is not justified a 
priori. However, such a value of $\Lambda_{QCD}$ leads to 
$C_{2} = -(0.05 \pm 0.06)\; \mbox{GeV}^{2}$ using Eq.(4), and
$C_{2} = -0.03 \pm 0.06 \; \mbox{GeV}^{2}$ using Eq.(7),
which are consistent with zero, and thus justifies the assumption
a posteriori.  All the above numerical calculations were performed
using an analytical chi-squared fit to the ALEPH data, and also by direct
integration of  these data , as they are available in bin-interval format
\cite{ALEPH}. The difference between the hadronic integrals in Eq.(3) from
these two procedures is less than 1 \%, which provides a reasonable
estimate of the error in integrating the data.
While this small difference translates into a larger
difference in the values of $C_{2}$, due to the cancellation between the
hadronic and the QCD integrals, the conclusion remains the same, i.e.
that given the present uncertainty in $\Lambda_{QCD}$ the ALEPH data imply
a value of $C_{2}$ consistent with zero. This conclusion is rather different
from that reached in a previous analysis \cite{CAD1}, along the same lines
as here but using instead the ARGUS tau-decay data \cite{ARGUS} to
determine the hadronic spectral functions. This discrepancy is due in
part to the much larger error bars of the ARGUS data. However, this does not
fully explain the differences, which are mostly due to different values
of the vector and axial-vector spectral functions at, and in the vicinity
of the respective resonances (rho- and $a_{1}$-mesons). This translates
into different areas under the hadronic spectral functions. For instance,
the ARGUS data saturated the first and second Weinberg sum rules reasonably
well \cite{CAD2}, while this is no longer the case for the
ALEPH data\cite{ALEPH}, \cite{CHSR}. \\

{\bf Acknowledgements}\\
One of us (KS) wishes to thank A.A. Pivovarov for helpful discussions.

\begin{center}
{\bf Figure Captions}
\end{center}
Figure 1. The right hand side of Eq.(9) as a function of the continuum
threshold $s_{0}$ in the vector channel (curve (a)),
the axial-vector channel (curve(b)), and the average
$C_{2} \equiv (C_{2V} + C_{2A})/2$ (curve (c)), all for 
$\Lambda_{QCD}= 300 \;\mbox{MeV}$. The expanded expression, Eq. (4), has
been used.

Figure 2. The average $C_{2}$ as a function of the continuum threshold
$s_{0}$ for $\Lambda_{QCD}= 300 \;\mbox{MeV}$ (curve (a)), and
$\Lambda_{QCD}= 400 \;\mbox{MeV}$ (curve (b)). The expanded expression,
Eq.(4), has been used.

Figure 3. The dependence of the average $C_{2}$ on $\Lambda_{QCD}$
for a typical value $s_{0} = 2 \;\mbox{GeV}^{2}$ at the center of the
stability region. The expanded expression, Eq.(4) has been used.
\newpage
\begin{figure}[h]
\begin{center}
\begin{picture}(400,300)(0,0)
\put(-40,360){\special{em:graph fig1.pcx}}
\end{picture}
\end{center}
\caption{ }
\end{figure}
\newpage
\begin{figure}[h]
\begin{center}
\begin{picture}(400,300)(0,0)
\put(-40,360){\special{em:graph fig2.pcx}}
\end{picture}
\end{center}
\caption{ }
\end{figure}
\newpage
\begin{figure}[h]
\begin{center}
\begin{picture}(400,300)(0,0)
\put(-40,360){\special{em:graph fig3.pcx}}
\end{picture}
\end{center}
\caption{ }
\end{figure}

\end{document}